\documentclass[]{aastex}
\usepackage{emulateapj5,epsfig}

\def\mpch{\mbox{$h^{-1}$Mpc}}
\def\Mpch{\mbox{$h^{-1}$Mpc}}
\def\Msun{\mbox{M$_\odot$}}

\def\Msunh{\mbox{$h^{-1}$M$_\odot$}}
\def\hmsun{\mbox{$h^{-1}$M$_\odot$}}

\def\Mvir{\mbox{$M_{\rm vir}$}}

\def\sigperp{\mbox{$\sigma_{\perp}$}}
\def\ltsima{$\; \buildrel < \over \sim \;$}
\def\lsim{\lower.5ex\hbox{\ltsima}}
\def\gtsima{$\; \buildrel > \over \sim \;$}
\def\gsim{\lower.5ex\hbox{\gtsima}}
\def\'#1{\ifx#1i{\accent"13\i}\else{\accent"13#1}\fi}

\newcommand{\lamfrac}{${\lambda_f^\prime}/{\lambda_i^\prime}$}
\newcommand{\lamf}{\lambda_f^\prime}
\newcommand{\lami}{\lambda_i^\prime}
\newcommand{\mfrac}{{M_f}/{M_i}}
\def\lt{\lambda^\prime}

\shortauthors{VITVITSKA et al}
\shorttitle{THE ORIGIN OF ANGULAR MOMENTUM IN DARK MATTER HALOS}
\begin{document}
\title{The Origin of Angular Momentum in Dark Matter Halos}
\author{Maya Vitvitska\altaffilmark{1},  
	Anatoly A. Klypin\altaffilmark{1},  
	Andrey V. Kravstov\altaffilmark{2},  
	Risa H.	Wechsler\altaffilmark{3,4},  
	Joel R. Primack\altaffilmark{3},  \&
	James S. Bullock\altaffilmark{5}
}  
\altaffiltext{1}
{Astronomy Department, New Mexico State University, Box 30001, Department 4500, Las Cruces, NM 88003}
\altaffiltext{2}
{Department of Astronomy \& Astrophysics, University of Chicago, 5640 South Ellis Ave., Chicago, IL 60637}
\altaffiltext{3}
{Department of Physics, University of California, Santa Cruz, CA 95064}
\altaffiltext{4}
{present address: Department of Physics, University of Michigan, Ann Arbor, MI 48109}
\altaffiltext{5}
{Department of Astronomy, The Ohio State University, 140 West 18th Ave.,
Columbus, OH 43210}

\begin{abstract}

We propose a new explanation for the origin of angular momentum in
galaxies and their dark halos, in which the halos obtain their spin
through the cumulative acquisition of angular momentum from satellite
accretion.  In our model, the build-up of angular momentum is a random
walk process associated with the mass assembly history of the halo's
major progenitor.  We assume no correlation between the angular
momenta of accreted objects.  The main role of tidal torques in this
approach is to produce the random tangential velocities of merging
satellites.  Using the extended Press-Schechter approximation, we
calculate the growth of mass, angular momentum, and spin parameter
$\lambda$ for many halos.  Our random walk model reproduces the key
features of the angular momentum of halos found in 
$\Lambda$CDM N-body simulations:
a log-normal distribution in $\lambda$ with an average of
$\langle\lambda\rangle \approx 0.045$ and dispersion
$\sigma_\lambda=0.56$, independent of mass and redshift.  The
evolution of the spin parameter in individual halos in this model is
quite different from the steady increase with time of angular momentum
in the tidal torque picture.  We find both in N-body simulations and
in our random walk model that the value of $\lambda$ changes
significantly with time for a halo's major progenitor.  It typically
has a sharp increase due to major mergers, and a steady decline during
periods of gradual accretion of small satellites.
The model predicts that on average the $\lambda$ of $\sim
10^{12}\Msun$ halos which had major mergers after redshift $z=3$
should be substantially larger than the $\lambda$ of those which did not.
Perhaps surprisingly, this suggests that halos that host later-forming
elliptical galaxies should rotate faster than halos of spiral
galaxies.

\end{abstract}



\section{Introduction}

Angular  momentum is  among the most  important quantities determining
the size and  shape of galaxies,  and yet a detailed  understanding of
its origins   remains a missing ingredient  in  the theory   of galaxy
formation.  \citet{Hoy}  was  apparently the first  astrophysicist who
discussed the source  of galaxy rotation in the  framework of a theory
of gravitational instability.  He explained galaxy rotation as arising
from gravitational coupling  with the surrounding matter.  Alternative
theories of the origin of galaxy rotation due to primordial turbulence
and vorticity were  also discussed at  that time  \citep{Weiz, Gamow},
but such  theories  were  subsequently ruled   out by  the   fact that
velocities not    arising  from  gravitation decay  in    an expanding
universe,  and also because of  improving
constraints on velocities derived from
cosmic microwave  background anisotropies.  

Hierarchical clustering of
cold dark matter (CDM) \citep{BFPR} is now widely believed to describe
the origin  of  structure in the   universe, and  in this  scheme  the
angular momentum of dark matter  halos --- and eventually the rotation
of  galaxies --- is  usually thought  to  be produced by gravitational
tidal   torques in the course  of  the growth of rotation-free initial
perturbations. This   is  a  well-developed   \citep{Peebles,  Dorosh,
White84}  and widely accepted theory, and  there  is little doubt that
the   tidal torques mechanism   is   connected with the generation  of
angular momentum.  To some  degree  the model  is supported  by N-body
cosmological simulations.  For example, if  one considers all the mass
which at $z=0$ ends up inside a given  halo, then the angular momentum
of that mass typically increases linearly with time at early stages of
the collapse, just as predicted by the tidal torque theory
\citep[e.g.,][]{BE87}.  At later times, again in agreement with
predictions, the growth in angular momentum slows down \citep{Suger}.

However, existing models and  approximations that try to implement 
the tidal torque scenario  do not agree  in detail with
the  results  of fully  nonlinear  N-body  simulations.  The predicted
angular momentum of halos   in  the tidal  torque model  is  typically
overestimated   by a factor   of $\sim 3$  compared  to the results of
simulations,  with a large scatter  of the same magnitude \citep{BE87,
Suger}. This over-prediction factor can be lessened if one assumes that
the angular momentum stops   its  growth earlier than  the  turnaround
moment \citep{Porciani2001} because of  nonlinear effects, but the
scatter still remains very large. The direction  of the spin parameter
predicted by the theory also  has large errors. \citet{Lee1} give  the
average error  in  the direction  of the spin  of about  57$^o$, which
agrees with \citet{Porciani2001}  who  find  a mean  misalignment   of
52$^o$ with a scatter $\sim 35^o$.  The  source of the errors is still
under  debate, but it seems  that the main effect  is due to nonlinear
effects which  are difficult to take into  account in the tidal torque
model \citep{Porciani2001}.

We also address another limitation of the tidal torque picture:
predictions for the spin parameter in this framework are for {\it all}
the matter which at $z=0$ is found in a given halo.  At high redshifts
this mass is of course not in a single halo.  In other words, the
tidal torque theory does not predict the rotation of any particular
progenitor of the halo in question. In this paper, we will pursue an
alternative approach, that of tracing the evolution of the angular
momentum of the most massive progenitor of the present-day halo.  This
alternative approach is much more useful for semi-analytic modeling of
galaxy evolution \citep[e.g.][]{SP99}.

The angular momentum can be expressed in terms of the dimensionless spin 
parameter, which is defined as
\begin{equation}
\lambda \equiv \frac{J { |E|^{1/2} }}{GM^{5/2}} \, .
\end{equation}
Here $J$ is the angular momentum, $E$ is the  total energy, and $M$ is the
mass of a halo.   The value of the  spin parameter roughly corresponds
to the ratio of the angular momentum of an  object  to that needed for
rotational   support  \citep[e.g.,][]{Pad}.    For  example,   a  spin
parameter of   $\lambda=0.05$ implies very  little systematic rotation
and  negligible rotational   support.   Typical  values of   the  spin
parameter  of individual  halos in  simulations  are $0.02$ to  $0.11$
\citep{BE87,  Ryden,     Warren,  Steinmetz,   Cole,    Gardner}.  The
distribution of spin parameters 
in N-body simulations is well described by the log normal distribution:
\begin{equation}
p(\lambda)d\lambda = {\frac{1}{\sigma_{\lambda}\sqrt{2 \pi}}}
{\exp\left(-\frac{\ln^2(\lambda/\bar{\lambda})}{2\sigma_{\lambda}^2}\right)
\frac{d\lambda}{\lambda}} .
\end{equation}
The parameters for the log-normal distribution were found to be $0.03
\leq \overline{\lambda} \leq 0.05$ and $0.5 \leq \sigma_{\lambda} \leq
0.7$ for standard CDM and various variants \citep[e.g.][]{Warren,
Gardner}.  For the $\Lambda$CDM cosmology with matter density
$\Omega_0=0.3$, $\Omega_\Lambda=0.7$, $h=0.7$, and $\sigma_8=1$, the
log-normal parameters were found to be $\overline{\lambda} =
0.042\pm0.006$ and $\sigma_{\lambda} = 0.50\pm0.04$ \citep{DTH}.
Note that the $\bar\lambda$ parameter of the log-normal distribution
(2) is not equal to the mean of $\lambda$; rather, 
$\langle \lambda \rangle \approx 1.078 \bar\lambda$ for
$\sigma_\lambda = 0.5-0.6$.

In order to study the evolution of the angular momentum of a dark
matter halo it is important to find correlations of the spin parameter
with other parameters of the halo and with its environment.
Correlation in the directions of spins of nearby halos or galaxies is
interesting as an indicator of the strength of large-scale correlations
and how much the next infalling satellite ''knows'' about the previous
one. Because the tidal torques are due to perturbations in the
gravitational potential, one naively expects that there are long range
correlations in spins of halos and, consequently, in the angular
momenta of galaxies. Indeed, there are claims that such correlations
exist in both observational data and in N-body simulations.  But the
correlations of galactic rotation axes are measured to be small.  If
$\hat L(\bf x)$ is a unit vector in the direction of the spin axis, then
\citet{PenLeeS2000} find that the correlation function of directions
$\langle \hat L({\bf x}) \cdot \hat L({\bf x} + {\bf r})\rangle$ is
less than 0.05 for distances $\lsim3~\Mpch$, and is even smaller at
larger distances. Even at small separations the effect is small and is
consistent with no correlations.  Results of N-body simulations
also indicate that spin correlations are weak at best \citep{BE87}.  
\citet{Lee1,
Lee2} found the spin-spin correlation function of halos is $\approx
0.05$. \citet{Porciani2001} found no correlation at distances
larger than $1\Mpch$ at late stages of evolution ($z=0$). These very
small correlations also suggest that material which is accreted by a
growing halo has very little memory: accretion is mostly random.

The spin parameter of halos in N-body simulations appears to be a very
stable statistic and has been shown to be independent of most physical
parameters. No dependence has been found on the cosmological model
\citep{BE87, Warren, Gardner, LemsonKauffmann}, on halo environment or
on halo mass \citep{LemsonKauffmann}. The only correlation that has
been shown to exist is with the time of the last major merger: halos
that experienced a recent merger have larger spin \citep{Gardner}.

In this paper we investigate the amount of angular momentum that
is brought in by the random uncorrelated accretion of satellites. Because
there is no systematic rotation of the incoming satellites, one might
naively expect that their contribution to the rotation of the
accreting object will be small.  But at the same time, each act of
accretion contributes a large orbital angular momentum.  Thus the halo
typically acquires a substantial rotation from each large merger,
which may later be diluted by many uncorrelated small mergers. The
outcome of the process was not obvious in advance, and to our surprise
we found that the resulting rotation is quite significant.

We begin, in \S 2, with an analysis of N-body simulations. 
There are two reasons for this.  First, we 
study the spin histories of several N-body
halos in order to provide 
a  comparative test for our model predictions.
Second, we need to characterize the distributions of velocities
and angular momenta of infalling satellites, because these are 
essential inputs for our model.
 
In \S 3 we present our random walk model for
the build-up of angular momentum in halos.  
Results of the random walk approximation are
presented in \S 4.  We discuss our results and the
general implications of our model in \S 5.

\section{Spin parameters, halos, and angular momenta of satellites in 
N-body simulations}

\subsection{Numerical simulations} 

We use three N-body simulations of a low-density flat $\Lambda$CDM
model with the following parameters: $\Omega_0=0.3$,
$\Omega_{\Lambda}=0.7$, and $h=0.7$. The simulations
were done with the Adaptive Refinement Tree (ART) code \citep{ART}.

The first simulation, \cite{KKBP00},
was done only for three halos, but the resolution
was extremely high: the mass per particle was $1.2\times 10^6\Msunh$
and the (formal) force resolution was $100h^{-1}$pc. Here we used
the power spectrum normalization $\sigma_8=0.9$.  At redshift
$z=0$ the three
halos had virial masses $(1.1-1.5)\times 10^{12}\Msunh$,
corresponding to $\sim 10^6$ particles per halo.

\begin{figure*}[t]
\begin{minipage}[t]{0.48\linewidth}
\centerline{\psfig{file=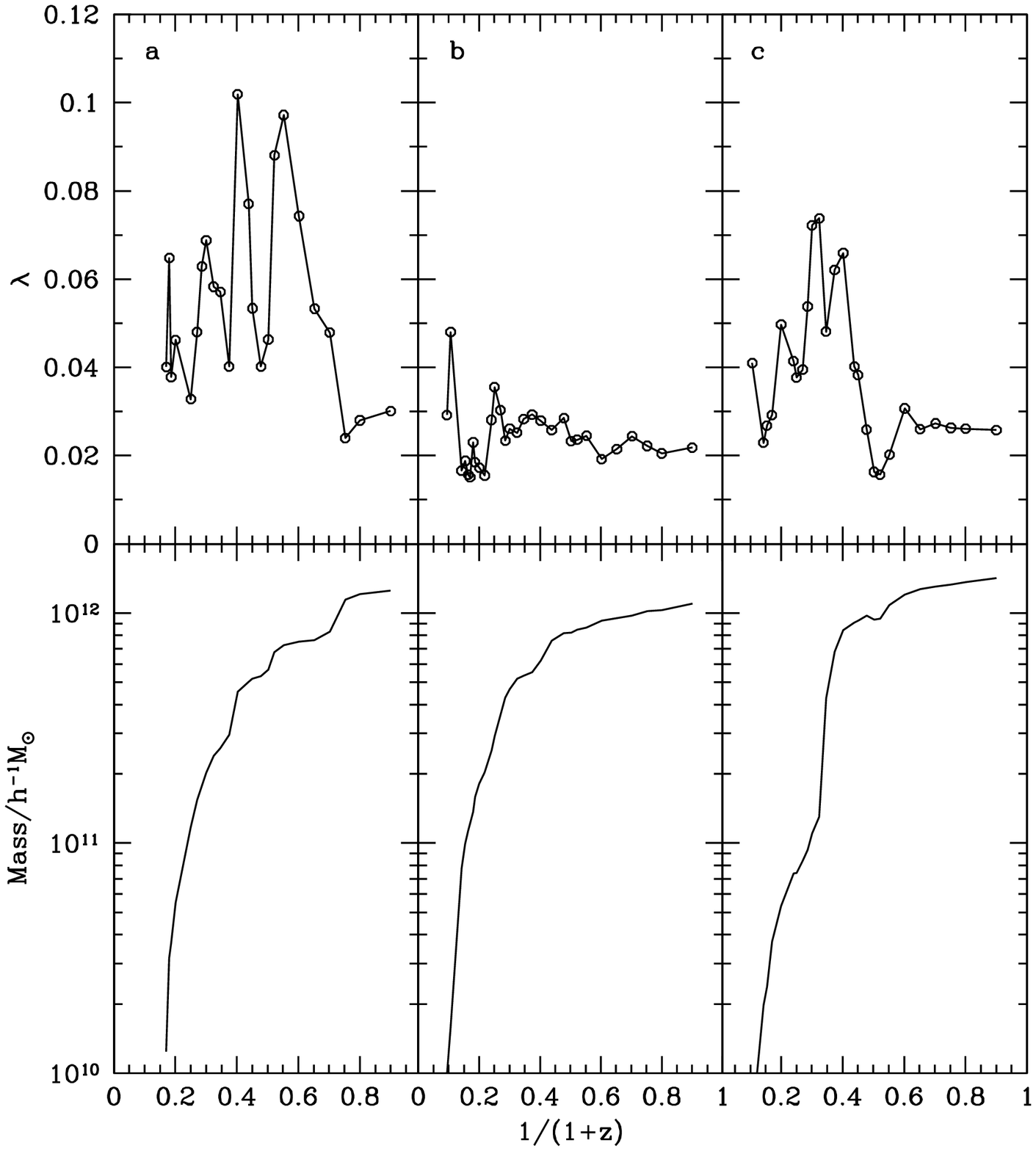, width=3.2in}}
\caption{\small Three examples of
evolution tracks of galaxy-size halos in N-body simulations. All halos
show fast mass growth at high redshifts. At that epoch their spin parameters
behaved very violently, but subsequently they mostly declined as the
halo masses grew.
}
\label{fig:Tracks}
\end{minipage} 
\hfill
\begin{minipage}[t]{0.48\linewidth}
\centerline{\psfig{file=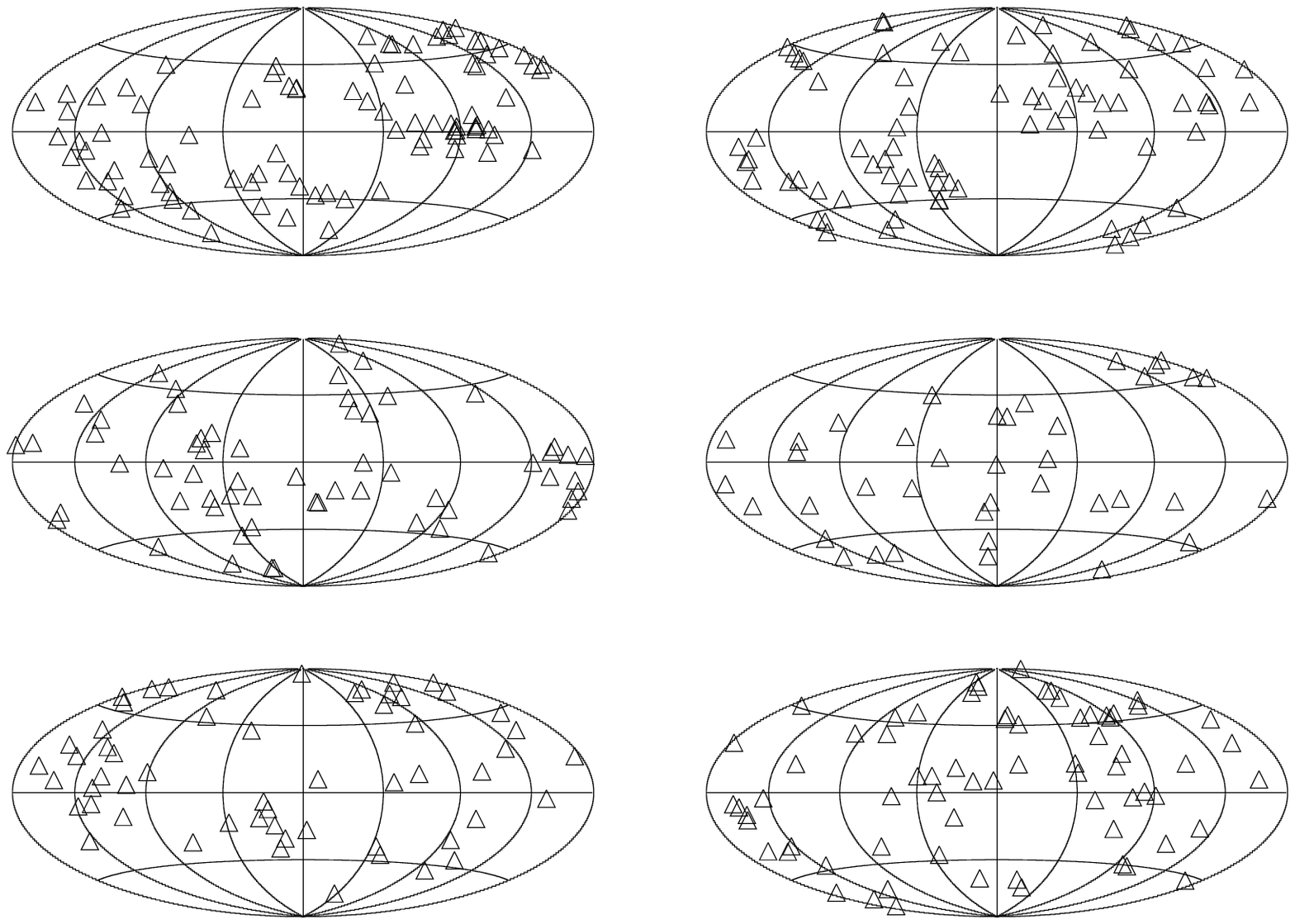, width=3.2in}}
\caption{\small The distribution of directions of the orbital
 angular momenta of
 accreted satellites relative to the direction of rotation of the
 central galaxy-sized halos. From top to bottom each row corresponds
 to halos $a$, $b$, and $c$. Left column is for halos at $z=1$; the
 right column is for $z=0$.  The angular momentum of the central halo
 corresponds to the northern pole (the top point in each panel).
 There is no clear correlation of the distribution of the directions
 of satellites either with redshift or with the direction of rotation
 of the central halo. }\label{fig:Directions}
\end{minipage} 
\end{figure*}

In order to find statistics of velocities and orbital angular 
momenta of satellites needed for the random walk model, we used two
additional simulations, which do not have as high resolution as the first
one but have many 
more halos. One simulation used $256^3$ dark matter
particles within a cubic box of comoving size $60 h^{-1}$Mpc with mass
per particle $m_p=1.1 \times 10^9 h^{-1} \Msunh$ and with a (formal)
force resolution $1.8 h^{-1}$kpc. For this simulation we used
$\sigma_8=1.0$.  This simulation has 15500 dark matter halos at
redshift $z=0$ with masses from $2\times 10^{10} \Msunh$ to $\sim
10^{15} \Msunh$. Another simulation used $512^3$ dark matter particles
within a cubic box of comoving size $80 h^{-1}$Mpc. The mass per
particle was $m_p=3.2 \times 10^8 h^{-1} \Msunh$. This simulation was run
only to $z=3$. It had 62,000 halos with mass larger than $1.3\times
10^{10} \Msunh$. This simulation was used to study statistics of large
mergers.

In our simulations the halos are identified by the
Bounded-Density-Maxima (BDM) algorithm \citep{KlypinHoltzman}. For
distinct halos (halos that are not inside larger halos) the
algorithm defines halos as spherical objects with average virial
overdensity. Halos inside larger halos are defined as gravitationally
bound lumps of dark matter.

\subsection{Evolution of halo spin parameters of major progenitors}

Figure \ref{fig:Tracks} shows the evolution of the mass and the spin
parameter of the major progenitors of our three $\sim 10^{12}\Msunh$
high resolution halos.  The spin parameter of the halos clearly
changes with time. Even for the most ''quiet'' halo $b$ the spin
parameter changed by a factor 1.5 since redshift 3. For the other two
halos the change was even larger. 
Note that there is no steady increase of the spin with
time; instead, there are large sudden changes in $\lambda$, which
clearly correlate with jumps in the masses of the halos mostly
associated with large mergers.
The jumps in $\lambda$ are of both signs: most large merger events are
associated with large spin increases, but halo $a$ had a relatively
large merger event at $z\approx 0.33$, which decreased its spin by
about a factor of two.  There is also a tendency of $\lambda$ to
decline during periods of gradual mass accumulation, which is
clearly observed in halos $b$ and $c$ at later stages 
of evolution.

The halos accrete satellites from all
directions. Figure~\ref{fig:Directions} shows directions of orbital
angular momenta of satellites which are being accreted  by the halos
at two redshifts. We show satellites that at a given redshift
are found in a spherical shell 
radius between 0.9 and 1.1 times
the virial radius. Only satellites that move toward the center are considered.
The distribution is not totally random: the satellite spins are
clustered. This clustering of orbital spins comes from real-space
clustering: there are larger satellites that are surrounded by
smaller ones. When a group of satellites falls in to the central halo,
they have almost the same angular momentum. 

Besides the clustering, there is no discernible
pattern in the directions of the satellites. For example, halo $b$ at
$z=0$ has a hole around the northern pole (direction of rotation of
the central halo). This would indicate that orbits of satellites are
far from the plane of rotation of the central halo. At the same time,
halo $c$ has an excess of satellites with the same direction of
orbital motion. 

\begin{figure*}[t]
\begin{minipage}[t]{0.48\linewidth}
\centerline{\psfig{file=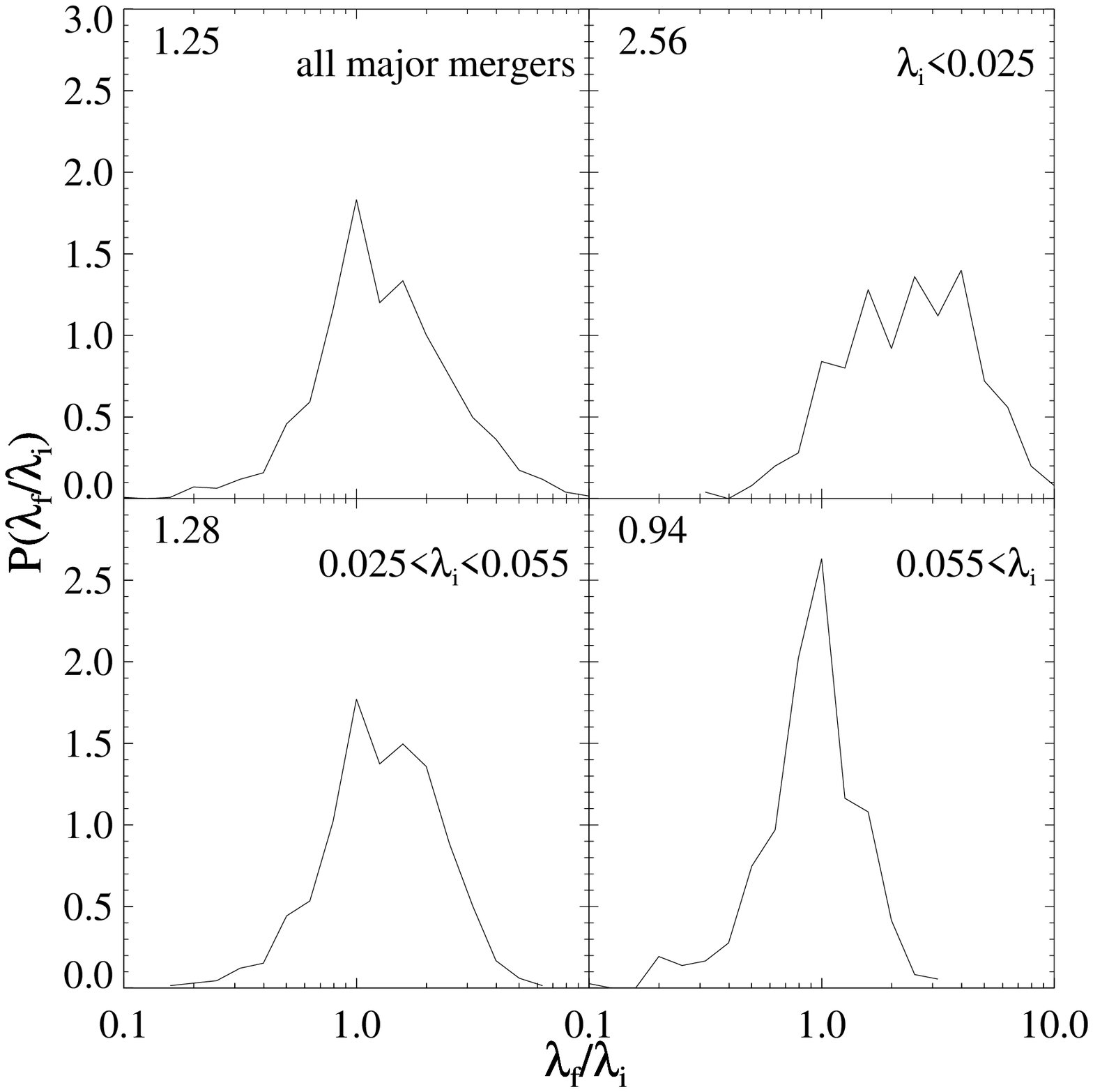, width=3.2in}}
\caption{\small Probability distribution of the ratio of final to
initial spin parameter \lamfrac\ between subsequent stored time-steps,
for halos with \Mvir$>5\times 10^{11}\hmsun$ at $z=0$ that have
undergone a major merger. Here $\lami$ is the spin parameter $\lt$ at the
last stored time-step before the merger, and $\lamf$ is the spin
parameter at the next stored time-step.  Panels represent: top left, all
halos; top right, $\lami<0.025$; bottom left, $0.025<\lami<0.055$;
bottom right, $\lami>0.055$.  The median value of \lamfrac\ is printed
in the upper left corner of each plot.  Halos with low initial spin parameters
typically have large increases in spin in a major merger.  
}\label{fig:lambda_mm}
\end{minipage} 
\hfill
\begin{minipage}[t]{0.48\linewidth}
\centerline{\psfig{file=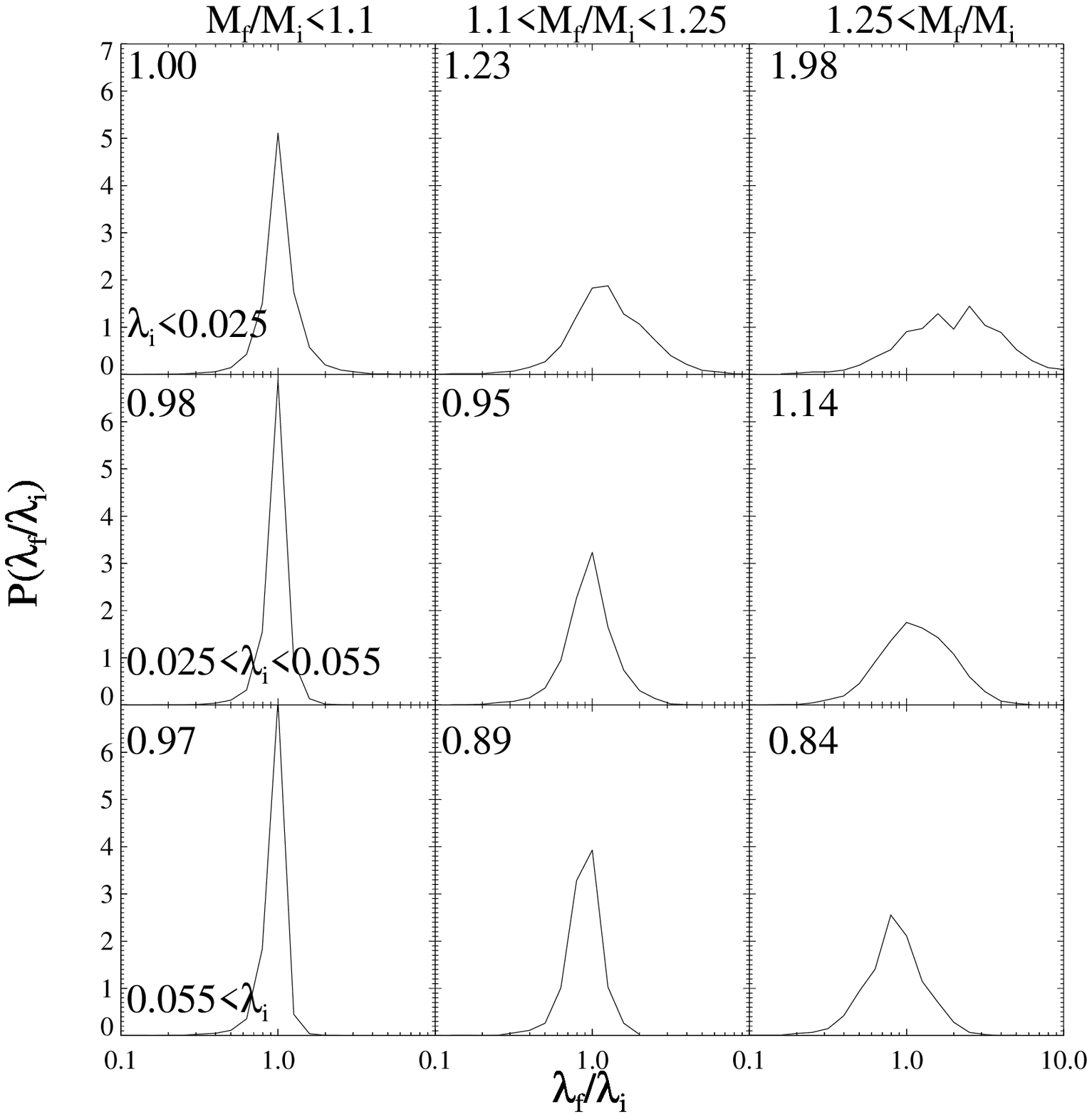, width=3.2in}}
\caption{\small 
Probability distribution of \lamfrac\, for halos with
\Mvir$>5\times10^{11}\hmsun$ at $z=0$.   
The various panels correspond to halos with different
initial spin parameters and different mass ratios.  The rows
correspond to increasing initial spin values from top to bottom:
$\lami < 0.025$, $0.025<\lami<0.055$, and $\lami>0.055$.  The columns
correspond to increasing mass ratios from left to right: $\mfrac<1.1$,
$1.1<\mfrac<1.25$, and $\mfrac>1.25$, where $M_f$ is the mass of the
halo in the later time-step and $M_i$ is the mass of that halo's most
massive progenitor in the earlier time-step. The median value of
\lamfrac\ is printed in the upper left corner of each panel.
}
\label{fig:lambda_change}
\end{minipage} 
\end{figure*}

These three halos were chosen to reside in a $\sim 10$ Mpc filament
bordering a large void. This is done to roughly mimic the environment
of the Local Group. The filament appeared at very high redshifts. The
three halos were always in the filament, growing by merging with smaller
halos.  A simple naive expectation is to see a lump of satellites in
panels in Figure~\ref{fig:Directions}. Its direction would indicate
the direction of the filament and the rotation of the halo relative to
the filament. We do not find this: there is no obvious large single
lump in the diagrams. The problem with the naive expectation is that
one imagines a thin tube with the halo inside it. Even qualitatively
this is a wrong picture. The filament is much wider (1-2~ Mpc) than
the virial radii of the three halos ($\sim 200$~kpc). When the satellites
eventually fall in to the halo, they come from different directions 
in the filament,
not from the main axis of the filament. Because of the finite
thickness of the filament, one may even expect to find
anti-correlation of orbital spins of satellites. Satellites moving
parallel to the axis bring opposite signs of the angular momentum when
they move above or below the axis.

Figure~\ref{fig:Directions} also shows that there is almost
no correlation with redshift for the same halo. Patterns are not
reproduced with time.
Statistical analysis of the directions confirms our visual
impression. The measured distribution is consistent with an uncorrelated
random distribution. The data still allow some degree of correlation
or anti-correlation because of the small number of satellites and
consequent statistical uncertainties. 
The total number of
satellites in our simulations was large, about 300 per
central halo at the end of evolution, yet the number of satellites in a 
shell around the virial radius was much smaller, about 40 for each
halo. With the total number of 120 satellites, statistical uncertainties
still allow correlation or anti-correlation on the level of 10-20\%.
Altogether, the evolution of the spin parameter and the
directions of accreted satellites indicate that the accumulation of
the angular momentum happens mostly in a random, uncorrelated fashion.

We used the second simulation with a large sample of halos to study
the correlations of jumps in $\lambda$ and in mass.  The analyses
presented here are based on a new structural catalog of halos from
this simulation, developed as part of the study of the relationship of
the structure of halos to their merging history in \citet{Wechsler}
and Wechsler et al.~(2002), where further details can be
found.  Briefly, all separate bound virialized halos with more than 20
particles are cataloged at each of 36 stored time-steps from this
simulation, and NFW fits \citep{NFW96,NFW97} are obtained for all halos
with more than 200 particles.  These catalogs were then compared in
order to determine the merging histories of all halos.  The statistics
presented here are for distinct halos, i.e.~those which are not
subhalos of another halo.

The simulations indicate that spin parameters of individual halos
change very substantially during the growth of halos: there are sharp
increases in $\lambda$ which correlate with periods of fast mass
increase, while periods of slow mass increase tend to correlate with
declines in the spin parameter.  Thus the features illustrated for
three halos in Figure \ref{fig:Tracks} are in fact quite general.  In
Figure \ref{fig:lambda_mm} we present statistics
of the change in the spin parameter in major mergers,
while Figure \ref{fig:lambda_change} includes these statistics
for a range of changes in the mass of the most massive progenitor.
In these figures, we actually plot the
changes not in $\lambda$ but rather in $\lt$, defined as \citep{DTH,Dekel}
\begin{equation}
\lambda^{'} = \frac{J}{{\sqrt 2} M_{\rm vir} V_{c} R_{\rm vir}},
\end{equation}
where $V_{c}^2=GM_{\rm vir}/R_{\rm vir}$ is the circular velocity at
the virial radius, $R_{\rm vir}$, for a halo with virial mass
$M_{\rm vir}$
(see \S 3.1 for more details).
The spin parameters $\lt$ and $\lambda$ are
approximately equal for typical NFW halos \citep{DTH}, but $\lt$ is
easier to measure in simulations.

For Figure \ref{fig:lambda_mm} we identified all halos with two
progenitors of mass ratio 
$m/M > 1/3$ in the previous stored
time-step, and investigate the distribution of \lamfrac\ in these
merging events.  On average, we find that $\lt$ increases by about
25\% in major merger events (Figure \ref{fig:lambda_mm}, upper left
panel), though there is a wide distribution.  Not surprisingly, the
change in $\lt$ is more severe for cases in which the initial $\lt$
was low; in these cases an incoming satellite is likely to increase
the spin regardless of its impact parameter.  Halos which have low
initial spin values can thus experience jumps of up to a factor of ten
in a single merging event.  When the initial spin is high, however,
merger events can serve to decrease the angular momentum if they come
in at opposite direction to the spin.  For very high initial values,
\lamfrac\ decreases slightly on average, occasionally as much as a
factor of five.  

Major mergers are clearly important since they can 
bring in a large amount
of angular momentum at once, but minor mergers also play a large role
in influencing halo spin.  In Figure \ref{fig:lambda_change}, we show
the distribution of \lamfrac\ for a variety of initial spin parameters
and changes in halo mass.  We see that the spin of halos with very low
initial spin values is likely to increase with any incoming material,
even a small amount.  For halos with intermediate initial spin values,
only major mergers are going to have a large effect on their spins,
and for halos with very large initial spin values, any incoming mass
is likely to decrease the spin.  Thus although major mergers produce
the largest jumps in $\lt$, we find that the entire mass accretion
history is important for its evolution.

\subsection{Velocities and angular momenta of satellites}

In this section we study the infall velocities of
satellites into larger halos in the simulations in order to provide
inputs to the random walk model.  We show that the distribution
of infall velocities can be well parameterized by a characteristic
Gaussian velocity dispersion and a radial anisotropy parameter.  In
addition, we find that the velocity anisotropy changes as a function
of the mass ratio of satellite to major progenitor, $m/M$: large
satellites tend have more radial orbits.  Since we would like to
include at least the basic trend with $m/M$ in our model, we make the
following simplified division: major mergers satisfy $m/M > 1/3$ and
minor mergers satisfy $m/M < 1/3$.  We characterize the velocity
ellipsoids of each of these populations separately.  
We sorted all halos by mass and selected pairs of halos starting with the
most massive ones.  Because we are interested in studying satellites
at the time of accretion, we selected only pairs separated by
approximately the virial radius $R_{vir}$ of the larger halo in each
pair. Three criteria were used: (1) the distance between
a pair of halos was required to be $(1 \pm 0.2)R_{\rm vir}$, (2) the relative
velocity of halos had to be less than $2V_c$, where $V_c$ is the
maximum circular velocity of the larger halo, and (3) the halos had to
have negative relative radial velocities (corresponding to inward
motion), i.e. the angle between the line from the smaller to the
larger halo and their relative velocity vector should be between $0^o$
and $90^o$.

The large sample of halos  in our $60 \mpch$  simulation allows us  to
study the statistics of   relative velocities and angular momenta   of
satellites falling onto larger halos. 
We did so by averaging results for expansion factors $a=0.972, 0.982,$ 
and $1.00$.
Numerical effects --- force and mass resolution and specifics of our halo
finder --- result in a rather complicated selection of mergers of
different mass ratios.  The limited force and mass resolution result
in destruction of most of satellites of small halos. This is the so called
overmerging problem.  Because of this problem the galaxy-size halos
in this simulation can not be used for studies of minor mergers, since
their satellites were not resolved. Small subhalos were resolved for
large groups and clusters, which had dozens or hundreds of them
\citep{Pedro}.  Thus, we studied minor mergers for
''host'' halos with masses larger than $10^{14}\Msunh$. Overall,
there were 543 satellites at radius $(1\pm 0.2)$ of the virial radius
of the ''host'' of which 301 were falling onto the ''host'' and the
rest were moving out. 
Merging with large satellites was studied mostly at high
redshifts. This was done using the $60 \mpch$ simulation at $z=1$ and
$z=3$. We also used the $80 \mpch$ simulation at $z=3$ to probe
statistics of satellites with different masses.

\begin{figure*}[t]
\begin{minipage}[t]{0.48\linewidth}
\centerline{\psfig{file=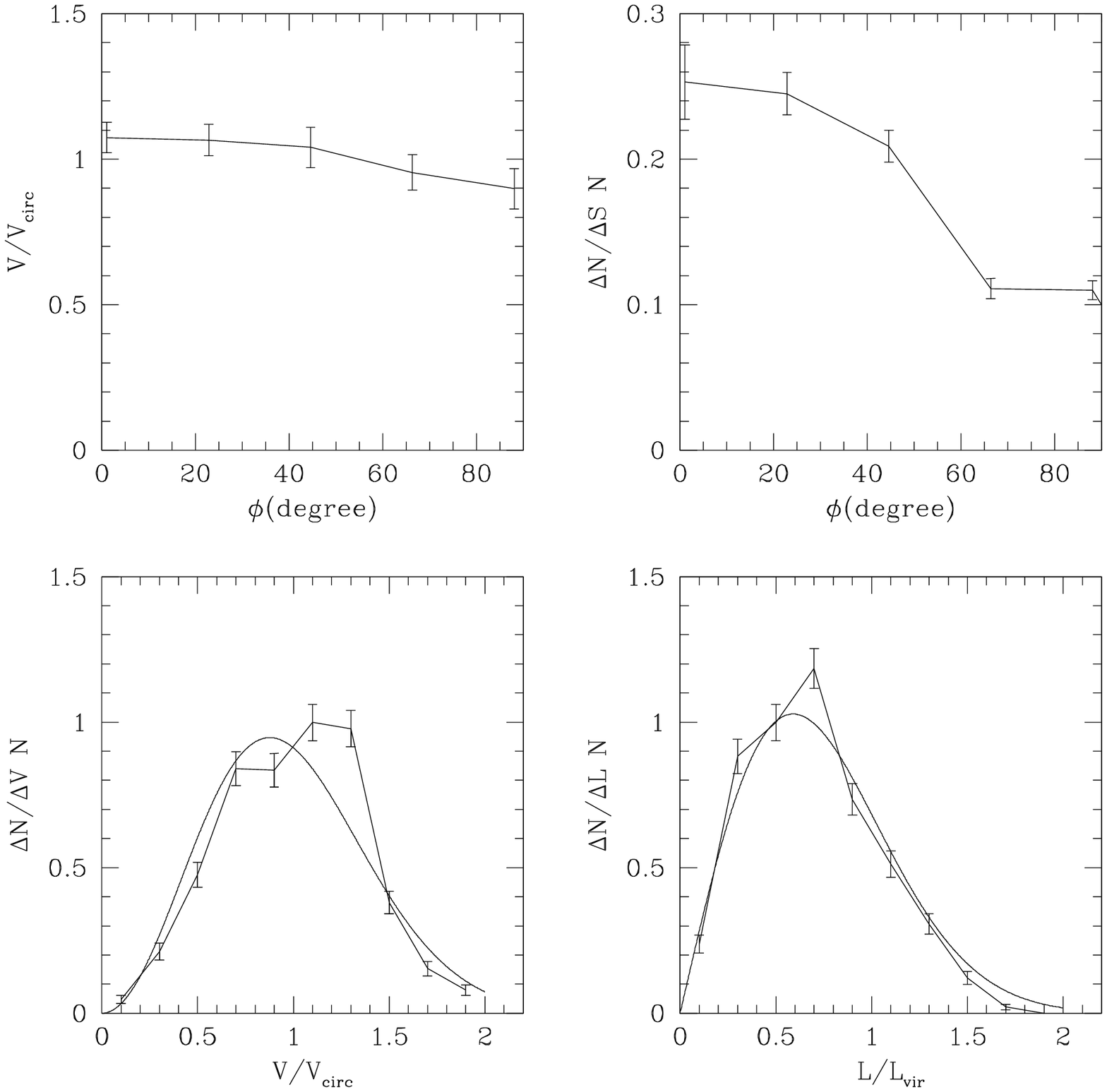, width=3.2in}}
\caption{\small The distributions of the angles, velocities, and
orbital angular momenta of infalling satellites 
for minor mergers identified in the simulation.
The angle $\phi$ represents the angle between the radius and
velocity vectors.  The top right panel shows the distribution of
$\phi$, and the top left panel shows the distribution with respect to
$\phi$ of the infall velocity in units of the virial circular velocity of the
larger halo.  The bottom panels show the distributions of the relative
velocities (left) and of the angular momenta (right).  The error bars
show the Poisson errors.  Smooth curves are the fits of the
distributions (see text for details).
}\label{fig:Catl60}
\end{minipage} 
\hfill
\begin{minipage}[t]{0.48\linewidth}
\centerline{\psfig{file=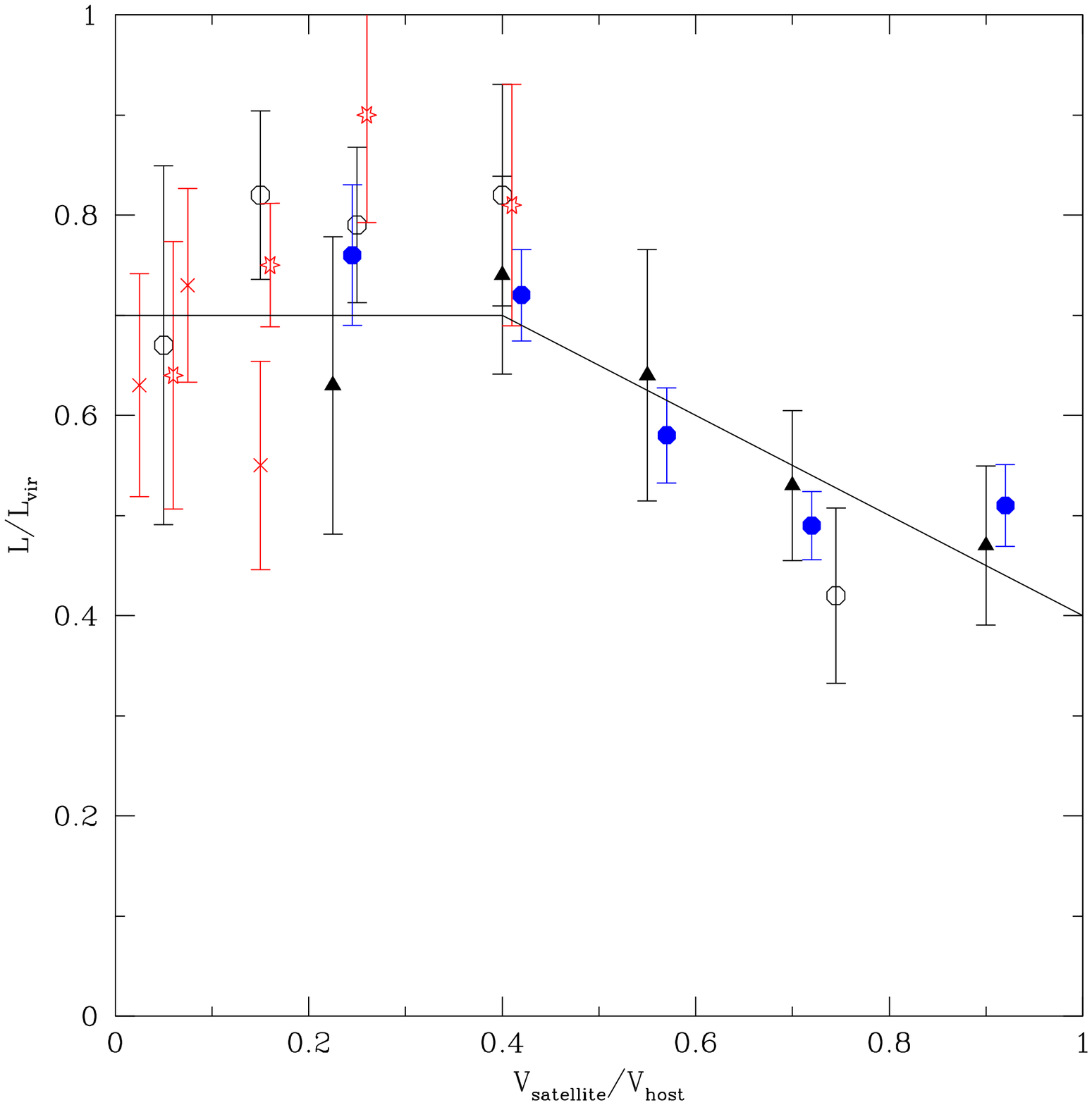, width=3.2in}}
\caption{\small The dependence of the specific angular momentum $L$ of a
  satellite  on the ratio of the satellite maximum circular
  velocity to the maximum circular velocity of the
  host halo. The specific angular momentum is measured in units
  of $L_{\rm vir} =R_{\rm vir}V_{\rm vir}$. The specific angular
  momentum of small satellites is larger than that of large
  ones; the line shows the fit of eq. (7).
  There is no dependence on the redshift. 
  Different markers show results for different simulations and
  redshifts. Stars, open circles, and triangles are for the $60\Mpch$
  simulation at redshifts $z=$0, 1, and 3 respectively. Crosses are
  for the three high resolution halos at $z=0$. Solid circles are for
  $z=3$ halos in the $80\Mpch$ simulation.}
\label{fig:sigma}
\end{minipage} 
\end{figure*}

Figure \ref{fig:Catl60} shows several statistics for the minor mergers
in  our catalog.   The upper  right  panel  shows the  distribution of
infall angles, $\phi$, of the smaller halos, where $\phi$ is the angle
between the  velocity    and radius vectors of   the   satellite.  The
distribution of relative  infall velocities in units  of $V_c$ of  the
larger halo is shown  in the lower  left, and the distribution of  the
relative specific angular momenta of satellites is shown in the bottom
right, plotted in  units of  $L_{\rm  vir}=V_{c} R_{\rm  vir}$ of  the
larger halo, where $ L \equiv \left|\bf{r} \times \bf {V}\right|$.
The error bars represent Poisson errors.

In the top left panel we show the distribution of the relative
velocity as a function of  $\phi$.
There is a small bias in the distribution of velocity
vs. angle, with higher infalling velocities corresponding to more
radial infall directions.  The distribution of the angles in the
top right panel shows that there are more infalling halos with radial
than with tangential motions.

In general, the magnitudes of the infall velocities are distributed
around the circular velocity of the massive progenitor (bottom left
panel). The distribution of the velocity can be fit with a 3D
Maxwell-Boltzmann distribution ($\sigma_v=0.62V_c$, 
$\langle  V^2\rangle=3\sigma_v^2$):
\begin{equation}
p(V)={\sqrt{\frac{2}{\pi}}}{\frac{V^2}{\sigma_v^3}}{\exp({-V^2/2\sigma_v^2})}.
\end{equation}
This fit provides a rough characterization of the 
distribution, 
but does not fully capture its nature.
This is likely because the the velocities are anisotropic,
and the Maxwell-Boltzmann distribution 
describes a system with isotropic velocities.
The distribution of satellite angular momenta (bottom right panel), 
or, equivalently, 
tangential  velocities,  is described reasonably well by a (2D)
Maxwell-Boltzmann shape: 
\begin{equation}
p(L)={\frac{L}{\sigma^2}}{\exp({-L^2/2\sigma_L^2})},
\end{equation}
with $\sigma_L=0.59 L_{\rm vir}$, $\langle L^2\rangle 
=2\sigma_L^2$.  A good fit is 
obtained because the tangential velocities are fairly isotropic.

These results suggest that the velocities of satellites are 
well-described by a 3D Gaussian distribution with a substantial
radial velocity anisotropy.  We further have checked
the distribution of each component of velocity and found them
to be roughly Gaussian.  
Based on this, it is  justified to summarize the 
velocity distribution of infalling satellites with
two parameters defining a velocity ellipsoid, 
one describing the velocity anisotropy,
\begin{equation}
\beta=1-\frac{\sigma_{\perp}^2}{2\sigma_r^2},
\label{eq:beta}
\end{equation}
and the other describing the radial Gaussian velocity
dispersion, $\sigma_r$.  For example, in the case of isotropic orbits
$\sigma_{\perp}^2=2\sigma_r^2$ and $\beta=0$, and in the case of nearly
radial orbits $\sigma_{\perp}^2 \ll 2\sigma_r^2$ and $\beta \approx
1$.  

From the 60 $h^{-1}$ Mpc 
numerical simulation, the velocity anisotropy and tangential
velocity dispersion were found to 
depend on the ratio $m/M$ of the mass of the satellite to that of the
host galaxy. For major mergers $\beta=0.8$ and
 $\sigperp=0.45V_c$, and for minor mergers $\beta=0.6$ and
 $\sigperp=0.71V_c$.  These values are in good agreement with
 \cite{Pedro}.  Note that major mergers are significantly more radial
 than minor mergers  and bring in a factor
 of 1.6 less specific angular momentum.  

We also studied the velocities of satellites in the simulation
of the three galaxy-size halos. The statistics of satellites in each halo
are much better in this simulation. We can also easily track the
evolution of each halo with redshift. The results are consistent with
what we found for the larger simulation.  For minor mergers the
average velocity anisotropy was $\beta =0.67$ at $z=0$ and $\beta
=0.53$ at $z=1$, however, the difference in this value is not statistically
significant, and on average the value remains unchanged ($\beta =0.6$).
The three halos in the first simulation have masses of $\sim
10^{12}\Msunh$. This is significantly smaller than
$10^{13}-10^{14}\Msunh$ in the case of the large-box second simulation
where, as we explained above,
most of the ''satellites'' were coming from group-size central
halos.  Thus, the results for the three halos suggest that there is
no (significant) trend of $\beta$ with the mass of the central halo.

The dependence of the angular momentum of infalling satellites on the
mass of the satellites is very important for accurate estimates of the
spin parameter $\lambda$. We use all available simulations to measure
this dependence. Because the mass of a satellite is not always a well
defined quantity (e.g., it depends on an uncertain truncation radius and
on details of removal of unbound particles), we prefer to use a more
reliable characteristic --- the maximum circular velocity.
Figure \ref{fig:sigma} shows the dependence of the
specific angular momentum of a satellite $L$ on the ratio of the
satellite maximum circular velocity $V_{\rm satellite}$ to the maximum
circular velocity of the host halo $V_{\rm host}$. 
Different markers show results for different simulations and
redshifts.  Error bars show $1\sigma$ shot noise estimates.

Figure \ref{fig:sigma} indicates that the specific angular momentum of
small satellites is larger than that of the large ones. It also shows
that there is no dependence of $L/L_{\rm sat}$ on the redshift.  The
same type of analysis shows that the
radial velocity dispersion does not depend on the mass of the
satellite: $\sigma_{r}/V_c=0.7\pm 0.1$. 
Thus, from eq. \ref{eq:beta}, 
velocity anisotropy $\beta$ varies
from $\beta =0.5$ for the small satellites to $\beta =0.8$ for the
major mergers.

We use the following approximation for the rms of the specific angular
momentum: 
\begin{equation}
 L/L_{\rm vir} = \left\{ \begin{array}{ll}
 0.7,                      &\mbox{if}\  V_{\rm sat}/V_{\rm host} < 0.4,\\
 0.9-0.5(\frac{V_{\rm sat}}{V_{\rm host}}), &\mbox{otherwise.} 
                           \end{array}\right.
\end{equation}
Figure \ref{fig:sigma}
 shows that this is a reasonable summary of the simulation
data.

\section{Random walk approximation}

The main idea of  our random walk model  is that halo angular momentum
is built  up  from the  summed  contributions of  uncorrelated orbital
angular momenta of accreted objects during the process of halo assembly.
The model needs several ingredients.  First,  we need to have the mass
accretion history for each halo (a merger tree) that tells us how many
satellites of what  mass  are accreted by  the  major progenitor as  a
function  of redshift.  For this, we  use the extended Press-Schechter
(EPS)   approximation  \citep{Press,Bower,BondEPS,LaceyCole}  and  the
EPS merger-tree method of \citet{SomKol}.~\footnote{This method describes
fairly   accurately the mass  accretion   history of major progenitors
in simulations
\citep{SLKD,Wechsler02}.}   
Second,  we  need  to know   the  position of  each
satellite at the  moment of accretion.  The position  is assumed to be
randomly distributed on a sphere whose radius  is the virial radius of
the main progenitor, with the main progenitor in the center.

Finally, we need to know the velocities of accreted satellites.
The three components of the velocities of each satellite
are assumed to be Gaussian, with a velocity anisotropy described by
eq. (6).  We use $\sigma_{r}=0.7 V_c$ regardless of the mass of the
satellite. The tangential velocity dispersion is assigned in the
following way. For a given satellite mass
we estimate its concentration parameter 
(see eqs. (16-17) below).  
The virial mass and
the concentration define the maximum circular velocity $V_{\rm sat}$.
We then use 
eqs. (6-7) to find the tangential velocity dispersion. A
realization drawn from  the Gaussian distributions gives the three
components of the velocity of the satellite.

\subsection{Mass, concentration, and spin parameter}

The model described above provides all of the information needed to
calculate the spins of an ensemble of halos.
The spin parameter $\lambda$ is determined by calculating the energy
for an NFW halo in virial equilibrium~\footnote{Our assumption of
virial equilibrium will likely break down just after a major merger
occurs.  We assume that equilibrium will be restored over a short
enough timescale that the general nature of spin evolution will not
depend sensitively on this detail.} and then using eq. (1).  For an
NFW halo, $\rho(r) \propto [(r/r_s)(1+r/r_s)^2]^{-1}$, the virial
concentration $c = R_{\rm vir}/r_s$, where the virial radius $R_{\rm
vir}$ is defined by the virial mass $M_{\rm vir}$ and by the
parameters of the cosmological model:
\begin{eqnarray}
M_{vir} & = & {\frac{4 \pi}{3}}{ \rho_{cr} \Omega_0 \delta_{th} r_{vir}^{3}}\\
r_{vir} & = & 443h^{-1} {\rm kpc \ } {\left(
\frac{M_{vir}/10^{11}h^{-1}M_{\odot}}{\Omega_0\delta_{th}}
\right)}^{1/3},
\end{eqnarray}
with \citep{BryanNorman}
\begin{eqnarray}
\delta_{th} &\simeq& (18\pi^2 + 82 x -39 x^2)/(1 + x), \\
 x &=& -(1 - \Omega_0)a^3 / [\Omega_0 + (1 - \Omega_0)a^3],
\end{eqnarray}
and $a=(1+z)^{-1}$.
  In the
$\Lambda$CDM model $\delta_{th} \sim 180$ at $a \ll 1$ and
$\delta_{th} \simeq 340$ at $a = 1$. 
The total energy of a halo truncated at the virial radius is
\begin{equation}
W = \frac{1}{2}\int  \rho(r) \phi(r)\, dr =
- {\frac{1}{2}}  {\frac{G M_{\rm vir}^{2}}{R_{\rm vir}}}  
{\frac{c g(c)}{f^2(c)}}.
\end{equation}
Here
\begin{equation}
f(c) = \ln (1 + c ) - \frac{c}{1 + c} 
\end{equation}
and 
\begin{equation}
g(c) =  1 - 2\frac{\ln(1 + c)}{1 + c} - \frac{1}{(1 + c )^{2}}.
\end{equation}
The spin is then
\begin{equation}
\lambda= {\frac{J}{2 V_{c} M_{\rm vir} R_{\rm vir}}} 
{\frac{[c g(c)]^{1/2}}{f(c)}}.
\end{equation}
For the median concentration
$c$, we use the results of \cite{Bullock}:
\begin{eqnarray}
c(M_{\rm vir},z) &=&\alpha(M_{\rm vir} )/ (1+z),\\ 
\alpha(M_{\rm vir})&\simeq& 
54.8-3.34\log({M_{\rm vir}/M_{\odot}}).
\end{eqnarray} 

In this paper we have assumed for simplicity that the halo mass and
redshift uniquely determines its concentration.  N-body simulations
indicate that there is a scatter in $c$ at fixed mass of the order of
35\% \citep{Bullock, Wechsler}.  
In principle, we could include
those deviations in our random walk model, but we decided to neglect
them because the deviations in $\lambda$ as a result of this
scatter are less than $10$ percent. 
\footnote{We also neglect the correlation between concentration and halo
merging history \citep{Wechsler,Wechsler02}.}

Thus, knowing the mass and the angular momentum of a halo one can find
the halo's concentration and its spin parameter.
\begin{figure*}[t]
\begin{minipage}[t]{0.48\linewidth}
\centerline{\psfig{file=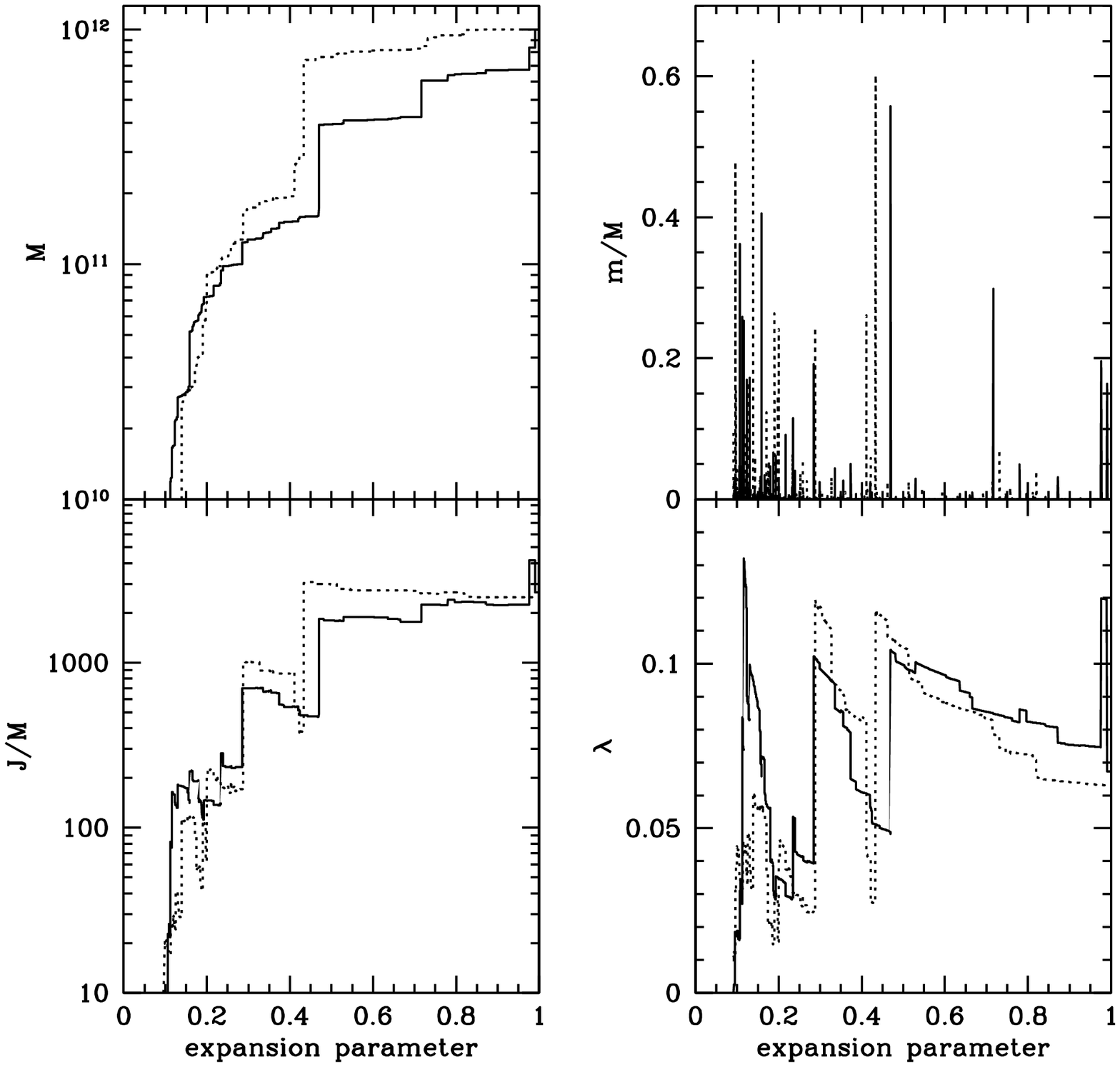, width=3.2in}}
\caption{\small The example of two typical evolution tracks (solid and
dashed lines) of the two dark matter halos.  On the left panels the
mass (top) and the specific angular momentum (bottom) are plotted as
functions of the expansion parameter $a=(1+z)^{-1}$. The right top
panel shows the mass ratio of the most massive merging satellite to
the main progenitor, where the case of major merger is defined as mass
ratio $m/M \ge 1/3$. The spin parameter as a function of the expansion
parameter is plotted in the right bottom panel.  
}\label{fig:track}
\end{minipage} 
\hfill
\begin{minipage}[t]{0.48\linewidth}
\centerline{\psfig{file=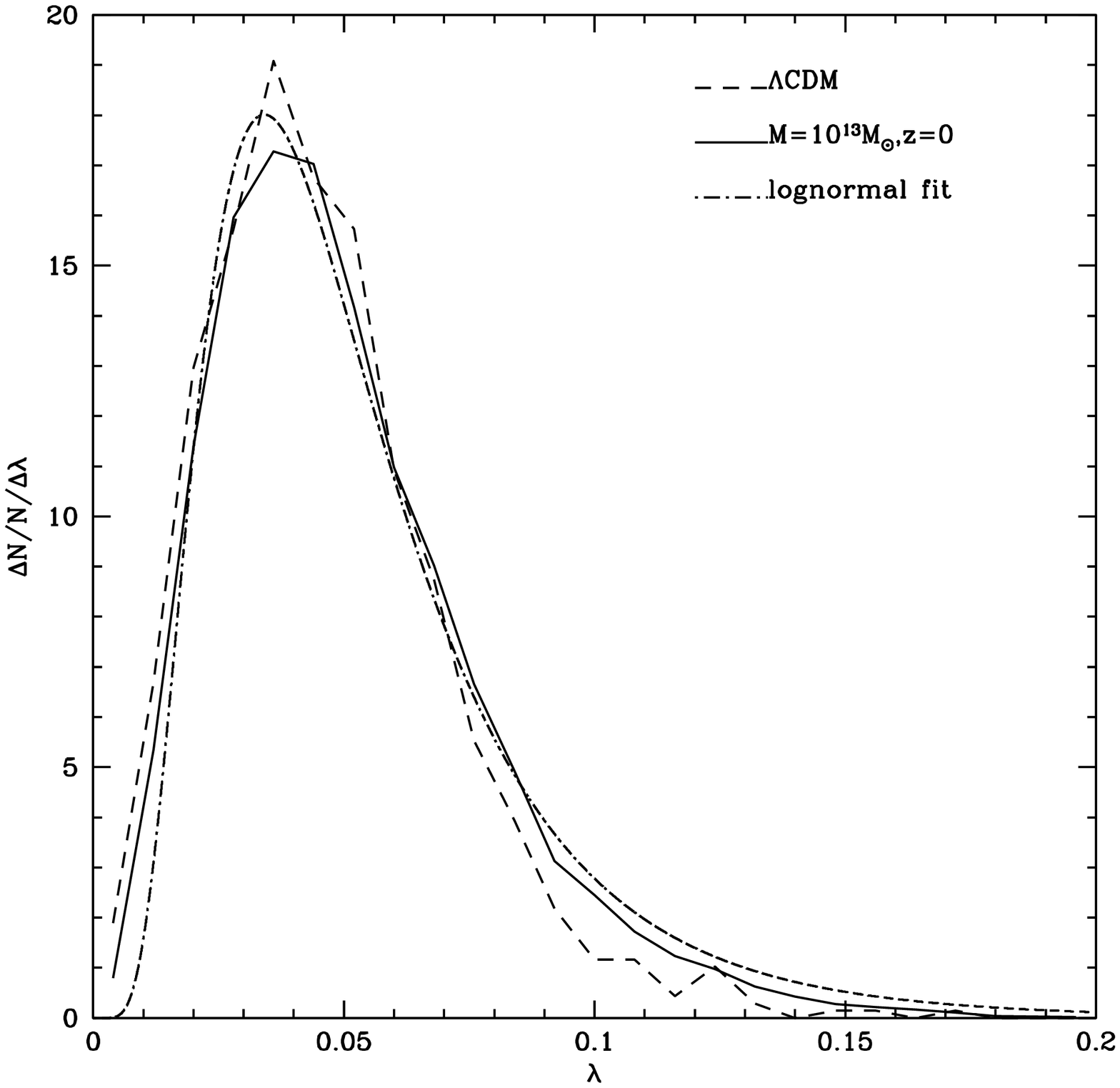, width=3.2in}}
\caption{\small Comparison of the distributions of the spin
parameter at redshift $z=0$. The solid line represents the
distribution of $\lambda$ derived using the random walk model.  The
distribution of $\lambda$ from our 60 $h^{-1}$ 
$\Lambda$CDM simulation is shown by
the dashed curve. The log-normal fit to the random walk model is
presented by the dot-dashed curve.  The parameters of the log-normal
distributions are in Table 1.}
\label{fig:mass_z}
\end{minipage} 
\end{figure*}

\subsection{Random walk model for the angular momentum}

Our main results are derived from two sets of Monte-Carlo merger trees
for two different final masses, $M_{\rm max}$.  Each set of runs
traced the merging history of the most massive progenitor back to
$z=10$ or until its mass fell below $M_{\rm min}$.  The first set had
$M_{\rm max}=10^{12}M_{\odot}$, $M_{\rm min}=10^8M_{\odot}$, and a
mass resolution (minimum mass of a satellite) of $M_{\rm
res}=10^6M_{\odot}$.  The second set of trees was done with $M_{\rm
max}=10^{15}M_{\odot}$, $M_{\rm min}=3 \times 10^8M_{\odot}$, and
$M_{\rm res}=10^8M_{\odot}$.  We used 300 mass tracks for each set and
each track was run 100 times with different realizations of satellite
velocities.  Unless otherwise stated, we use these two sets of merger
trees.
  
The evolution of the mass of two typical $10^{12}M_{\odot}$ tracks as
a function of the expansion parameter is shown on the top left panel
in Figure \ref{fig:track}. The right top panel shows the ratio of the
mass of the biggest satellite to the mass of the main progenitor.  The
mass of the main progenitor does not change smoothly, because we
assume that all new incoming mass instantly merges with the main
progenitor.

When the mass of the merging object is significant, there are often
dramatic changes in the specific angular momentum (bottom left panel)
as well as in the spin parameter (bottom right panel).  This also was
seen for simulated halos in \S2.  Of course, major mergers do not
always increase the spin parameter; depending on the velocity
orientation, such mergers can also decrease the halo's spin. In spite
of all these changes in the spins of individual halos, especially in
the past when the number of major mergers was significant, we will
show below that the {\it distribution} of the spin parameter doesn't
depend on time or on the mass of the main progenitor.

These examples show that the angular momentum doesn't increase
linearly.
Whether there are increases or
decreases 
in any given merging event 
depends on the mass of the satellite, its direction of
motion, and its velocity.  One can also see that at the later stages
of the evolution of a dark matter halo, when the mass of the main
progenitor doesn't change much, the angular momentum doesn't have
dramatic changes either.  This behavior is very similar to that found
in the N-body simulations, although the comparison cannot be made
precise because of the finite differences (up to $\sim0.5$ Gyr)
between the stored time-steps in the simulation compared to the
instantaneous approximation used in our random walk model.

\section{Results}

\subsection{Comparison with numerical simulations}

We begin by illustrating that our random walk approximation
naturally produces a log-normal distribution
in $\lambda$.  
Figure \ref{fig:mass_z} shows a comparison between N-body results and
our random walk model of spin acquisition.  The simulation corresponds
to $\Lambda$CDM ($\Omega_0=0.3$, $h=0.7$, $\sigma_8=1$). For the
random walk model we used 200 Monte Carlo merger trees with a final
($z=0$) mass of the main progenitor $M=10^{13}M_{\odot}$. Each mass
track was run 100 times with different realizations of satellite
velocities.  The distribution of the spin parameter from the N-body
simulation is shown by dashed line while the distribution from the
random walk model is shown by the solid curve. As one can see, there
is a reasonably good agreement between the N-body results and the
random walk model.  To demonstrate that these distributions are well
described by the log-normal function we also show a log-normal fit to
the random walk data (the dashed-dotted curve).  Parameters of the fit
are given in Table 1, along with spin parameters measured from N-body
simulations.

\begin{figure*}[t]
\begin{minipage}[t]{0.48\linewidth}
\centerline{\psfig{file=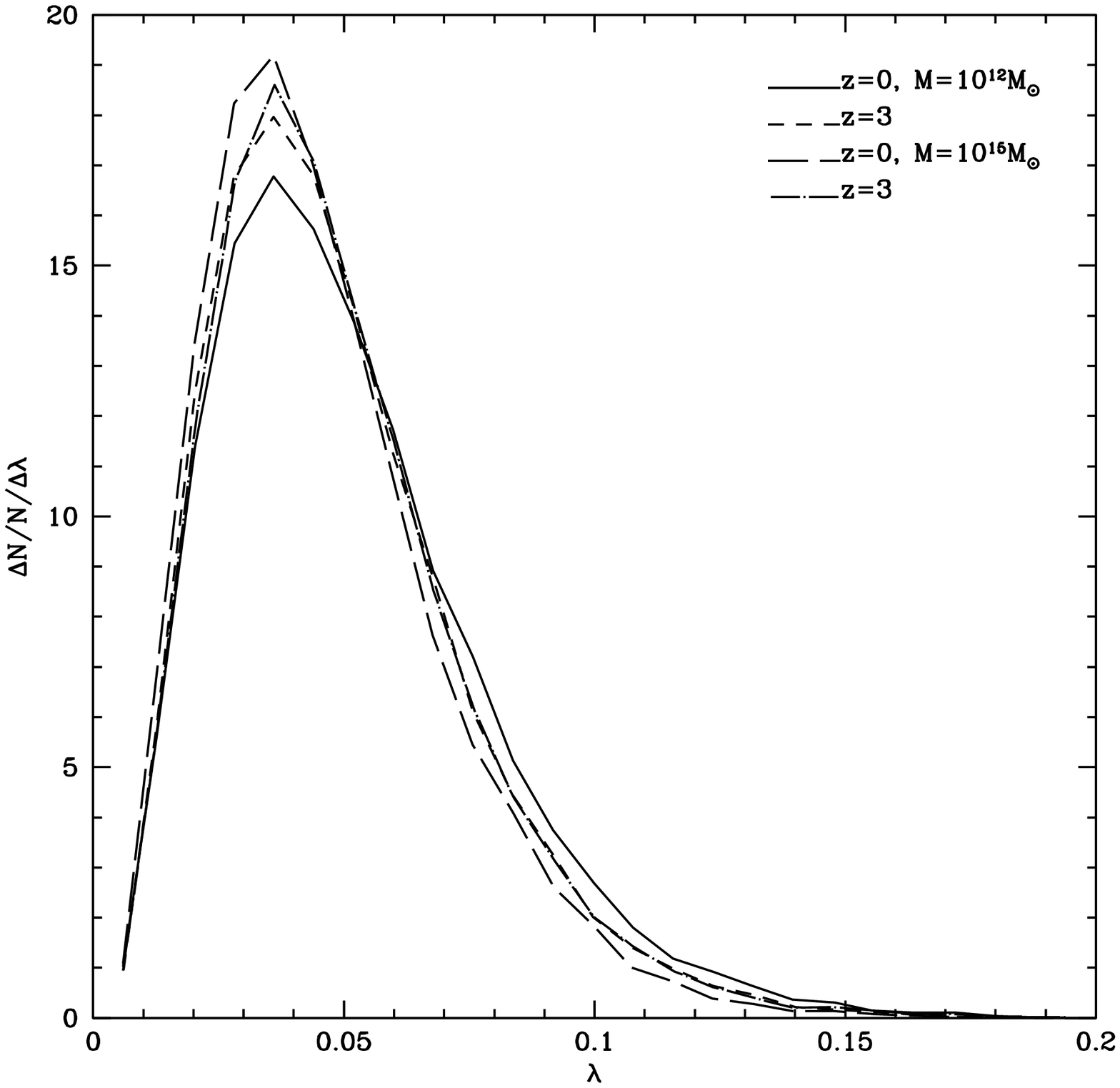, width=3.5in}}
\caption{\small 
The distribution of the spin parameter for 
halos at z=0 and z=3, for halos whose main progenitor 
has mass $10^{12}M_{\odot}$ or
$10^{15}M_{\odot}$ at redshift $z=0$.
The spin distribution has rather weak dependence
on mass and redshift.
 }
\label{fig:Redshift}
\end{minipage} 
\hfill
\begin{minipage}[t]{0.48\linewidth}
\centerline{\psfig{file=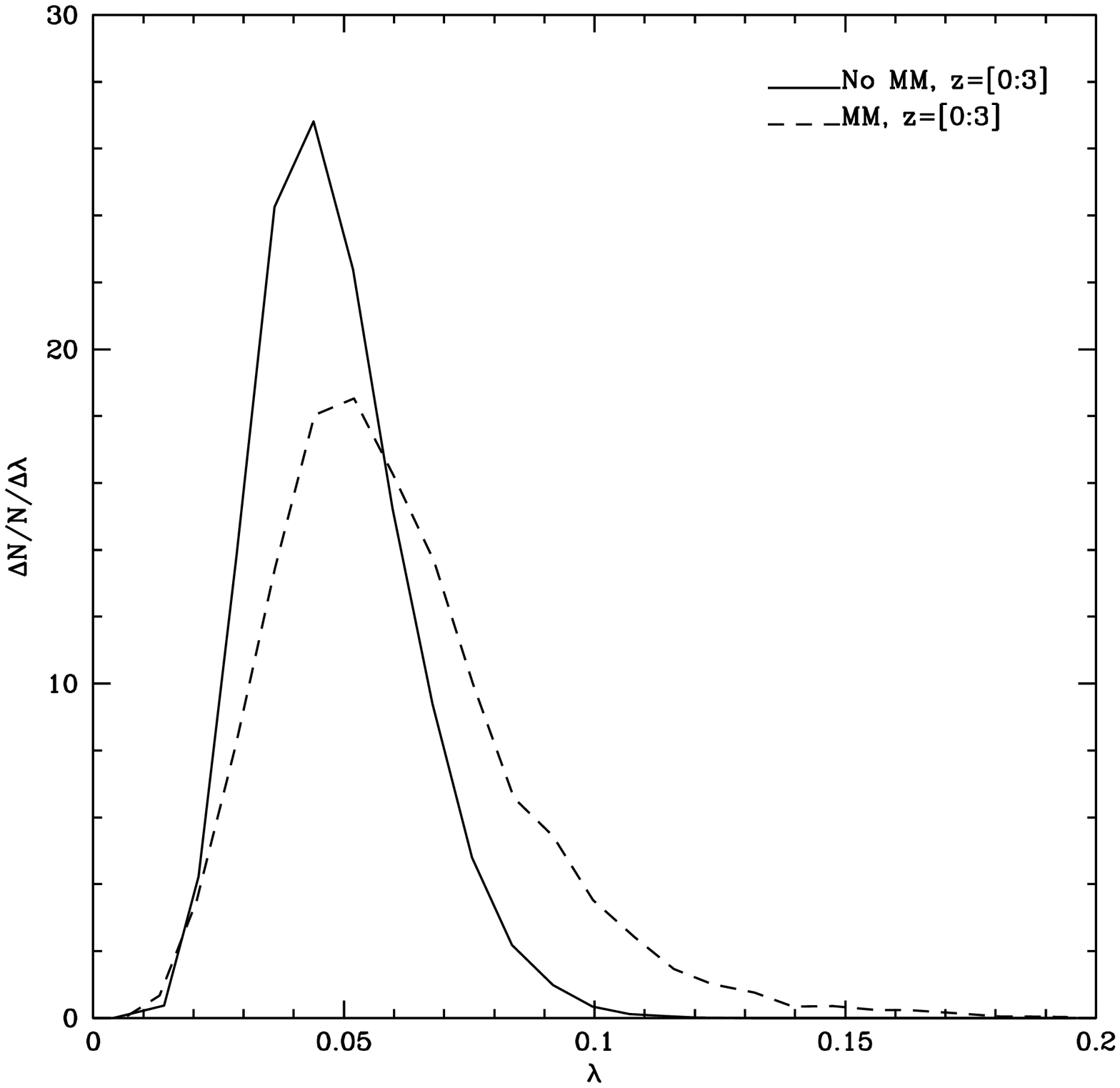, width=3.2in}}
\caption{\small 
The distribution of the spin parameter for dark matter
halos that had (dashed line) and didn't have (solid line) a major
merger event since redshift $z=3$. Final mass of the main progenitor
at $z=0$ is $10^{12}M_{\odot}$. The halos that had major mergers
during the last $\approx 11$ Gyrs on average have larger (by 25\%) spin
parameter than those which did not experience a major merger.
}\label{fig:MMeger}
\end{minipage} 
\end{figure*}

\subsection{Dependence of the spin parameter on mass and redshift }

As we discussed in \S1, N-body simulations indicate that the $\lambda$
distribution does not vary as a function of  redshift or halo mass.
Here we test our model against these results.
We analyzed two different sets of merger trees: one with
major progenitor mass $M=10^{12} M_{\odot}$ at redshift $z=0$, and the
other with mass $M=10^{15} M_{\odot}$ (see \S3). For both cases the
distributions of the spin parameters were examined at two different
redshifts: $z=0$ and $z=3$. These are shown in Figure
\ref{fig:Redshift}. The spin parameter distributions derived from the
random walk model show no correlation with either the redshift or the
halo mass; the distributions overlap and each has mean value $\langle
\lambda \rangle = 0.045$.

\subsection{Major mergers and the spin parameter}

As was just shown, the spin parameter distribution doesn't depend on
the redshift or on the mass.  But Figure \ref{fig:track} indicates
that $\lambda$ of individual halos in the random walk model is
sensitive to whether they have recently had major mergers, as we found
in N-body simulations (Figures 1, 3). In this subsection we study the
influence of these types of mergers on the distributions of the spin
parameter.

We study the effect of major mergers on the distribution of the spin
parameter by splitting all halos into two groups: the halos that
didn't have any major mergers in the past roughly 11 billion years ($0
\le z \le 3$), and the halos that did have a major merger during this
period. Since major mergers destroy galactic disks and produce
spheroidal stellar systems \citep[see e.g.][]{Barnes}, while mergers
with mass ratio less than 1/10 probably do not \citep[e.g.][]{Walker},
and stellar disks are typically up to about 10 Gyr old, only the halos
that did not have a major 
merger since $z\sim3$ could host mature spiral
galaxies.  The distributions of $\lambda$ for these two groups of
halos are shown in Figure \ref{fig:MMeger}. The mean value of the spin
parameter is $0.0465$ and $0.0569$ for no major mergers and major
mergers since $z=3$ respectively. in other words, recent major mergers
increase the spin parameter almost by 25\%. Thus, we find the same
result that was found in the N-body simulations: the spin parameters
are sensitive to major mergers.

\begin{table*}
\label{tab}
\begin{center}
\caption{Parameters of log-normal approximations for the spin distribution in a $\Lambda$CDM Model.}
\begin{tabular}{lccc}
\hline
\hline
Model & {$\overline \lambda$}   & {$\sigma_{\lambda}$}    & Comments\\
\hline
Numerical simulations: & & &\\
\qquad all halos at $z=0$       & 0.0445 & 0.563 & Figure \ref{fig:mass_z}\\
\qquad halos that had no major merger since $z=3$  & 0.034\\
\qquad halos that had a major merger since $z=3$   & 0.044 \\
Random walk model: & & & \\
\qquad halos with $M=10^{13}M_{\odot}$, $z=0$ & 0.0465 & 0.556
     & Figure \ref{fig:mass_z}\\
\qquad halos with $M=10^{12}M_{\odot}$ $\&$ $10^{15}M_{\odot}$,
  $z=0$ $\&$ $z=3$ 
 & 0.0450 &  0.560 & Figure \ref{fig:Redshift}\,(mean values)\\ 
\qquad halos that had no major merger since $z=3$  & 0.0465 & 0.333 &
  Figure \ref{fig:MMeger}\\
\qquad halos that had a major merger since $z=3$   & 0.0569 & 0.407 &
 Figure \ref{fig:MMeger}\\
\hline
\end{tabular}

\end{center}
\end{table*}

\section{Discussion}

We have  presented a new model for  the origin of the angular momentum
of dark matter halos in which angular momentum is built up in a random
walk fashion by mass accretion events.   The evolution of halo angular
momentum in  this picture is  quite different from that  inferred from
the  standard  tidal torques argument,  in  which the angular momentum
grows steadily at early times and its growth 
flattens out at late times.  Unlike
this standard picture,  we do not track the  angular momentum  of {\it
all} matter that  is in the present-day halo;  we track only the major
progenitor of the halo. 
The evolution of the spin parameter $\lambda$ of the major progenitor
is a random process in which the the value of $\lambda$ can vary by
factors of a few over its history, with a tendency for sharp increases
due to major mergers and steady decline during periods of gradual
accretion of small satellites.  These general properties of evolution,
including large fluctuations in $\lambda$, are confirmed by N-body
simulations.

Our model predicts that the distribution of the spin parameter is well
approximated by the log-normal distribution, with mean and dispersion
that do not depend on redshift or on the mass of the main progenitor.
The model predicts that on average the spin parameters of halos that
had major mergers after redshift $z=3$ should be considerably larger
than the spin parameters of those halos that did not.  Perhaps
surprisingly, this implies that halos which host later-forming
elliptical galaxies should rotate faster than halos of mature spiral
galaxies.

There may actually be observational evidence of this.  Planetary
nebulae in the two elliptical galaxies NGC 5128 (radio continuum
source Cen A) \citep{Hui95} and NGC 1316 (Fornax A)
\citep{Arnaboldi98} are rotating rapidly about their minor axes at
large radii.  The metal-rich globular clusters of NGC 5128 show
rotation similar to that of the planetary nebulae.  NGC 5128 is
regarded as a prototypical remnant of the merger of two disks, and NGC
1316 is also considered a likely remnant of a fairly recent merger.
From numerical simulations \citep{Barnes92,Hern93} of mergers between
disk galaxies with bulges, one indeed expects the angular momentum of
the remnant to be concentrated in the outer regions.  Similar studies
of the planetary nebulae of other elliptical galaxies such as NGC 4697
detect less evidence of strong rotation \citep{Mendez01}, and stellar
spectra of only a relatively small fraction of elliptical galaxies
show strong rotation at large radii (with $v_r/\sigma \gsim 1/2$),
e.g. NGC 1395 and NGC 1604 \citep{Franx89}.  The globular clusters of
M87 and a few other giant elliptical galaxies are also rotating
rapidly at large radii \citep{KPGebhardt,Cohen,Cote}.  Galaxies like
the cD M87 are thought to form from multiple mergers, and simulations
of this process predict that while velocity dispersion should dominate
in the central regions, rotation can grow to $v_r/\sigma \sim 1$ in
outer regions \citep{WeilHern}.  Radial velocity measurements for the
planetary nebulae and globular cluster systems of nearby elliptical
galaxies should become much more feasible with the advent of
multi-object spectrographs on 8-meter class telescopes.  Once data are
available on a larger number of such systems, it will be interesting
to see whether the statistics are consistent with predictions of the
random walk model for the origin of galaxy rotation that we have
proposed here.

Implementation of  the random  walk   model is a  rather   complicated
procedure if  one wants to reproduce N-body   results with an accuracy
better  than about 20   percent.    It  requires  measurements of    many
parameters of halos   and satellites  (e.g., halo concentrations   for
different  redshifts and   merging  trees).  The  most important  and,
unfortunately, the  most uncertain are  the parameters of the velocity
ellipsoid  of   accreted   satellites for different    masses   of the
satellites.   In principle, these parameters  could vary as a function
of redshift as well.  Estimates for the numerous minor  mergers seem to be
reasonably  reliable (see \S2.3).  Major 
mergers are rarer and   therefore  the  estimates are  more   uncertain.
Generally we find   that the trajectories  of mergers  with high $m/M$
ratios are  more radial than their low  $m/M$ counterparts,  but it is
difficult  to   measure   the parameters of    the velocity  ellipsoid
accurately.  

The model assumes that there is no net angular momentum of infalling
satellites and that the satellites are accreted in random
fashion. This picture is supported by results of N-body
simulations (\S2.2). Yet the results allow some degree of either correlation
or anti-correlation of orbital spins of the satellites. The allowed
effects are rather small (less than 10-20\%), but they still
may be important for accurate predictions of the angular momenta of
halos.

Our colleagues  \citet{Maller} have  investigated a simplified version
of what we call the  random walk model for the  origin of the  angular
momentum  of dark matter halos,  in  comparison with a  version of the
tidal torques model more sophisticated than the standard one discussed
briefly in \S1 of  the present paper.   They find that both models can
reproduce the log-normal distribution of halo  spin parameters seen in
simulations, with  appropriate tuning of the  model parameters.  Their
work is thus complementary to that presented here.

We expect  that our random  walk model for the  origin of halo angular
momentum  through the   accretion of satellites   will be  useful both
conceptually and practically.  In addition to giving a much better fit
to the  distribution  of halo  spin  parameters than the tidal  torque
theory,  it  also accords better  with  the hierarchical nature of the
process by which halos grow in the cold  dark matter model.  Moreover,
because it is  based on the   same extended Press-Schechter  formalism
used  in semi-analytic approaches   and  because it   is so simple  to
implement, we expect  that it will prove  invaluable  in improving the
treatment of angular momentum in modeling  the formation and evolution
of  galaxies.  For these same reasons,  it will  be worthwhile to test
and improve the random walk model.

We conclude with some comments about the implications of our model for
disk galaxy formation and the angular momentum problem in disks.  This
random walk picture   is  somewhat at  odds  with the   formalism made
popular by Fall  \& Efstathiou (1980)  in which  the gas is  initially
well mixed within a smoothly  rotating halo, and subsequently falls in
to form an angular momentum supported disk.\footnote{This picture was
further developed by \citet{BFFP}, \citet{FPBF}, and \citet{MMW98}, 
who took into account
the response of the dark halo to the infalling baryons.}
In our random walk model, the halo
angular momentum obtains its  smoothly varying distribution through  a
series  of clumpy merger  events and  subsequent relaxation.  Unlike a
picture in which tidal  torques determine the  spin of a halo  at very
early times, when the gas and dark matter was well-mixed, our scenario
would suggest  that the angular momentum distribution  of gas could be
considerably  different than that   of the  dark matter.  The  merging
satellite
galaxies are likely to be strongly affected by  the tidal field of the
central object,  and their gas is likely  to shock against  the gas in
and around  the host.  Since  it  is known  that the  angular momentum
distribution  in dark halos is  quite  unlike that  in disk galaxies
\citep{DTH,  Bosch2}, generally with an   excess of low-spin material,
our  random  walk picture  might   provide an interesting  avenue  for
exploring   solutions  of  
this angular momentum distribution problem and also of
the  well-known   angular  momentum  overcooling problem
(e.g. Navarro \& Steinmetz 1997).  Work in this direction is
under way.

\section*{Acknowledgments}
We acknowledge support from NASA and NSF grants at
NMSU, Ohio State, and UCSC, and we thank Avishai Dekel, Tsafrir
Kolatt, Ari Maller, and Rachel Somerville for helpful discussions.
A.K. is grateful to the Institute of Astronomy at Cambridge for
hospitality and financial support during his extended visit to 
Cambridge University.  A.V.K. was supported by NASA through a Hubble
Fellowship grant from the Space Telescope Science Institute, which is
operated by the Association of Universities for Research in Astronomy,
Inc., under NASA contract NAS5-26555


\begin{thebibliography}{47}
\expandafter\ifx\csname natexlab\endcsname\relax\def\natexlab#1{#1}\fi
\bibitem[Arnaboldi et al.~(1998)]{Arnaboldi98} Arnaboldi, M., et
 al. 1998, \apj, 507, 759
\bibitem[Barnes \& Efstathiou (1987)]{BE87} Barnes, J.E., \&
 Efstathiou, G. 1987, \apj, 319, 575 
\bibitem[Barnes (1992)]{Barnes92} Barnes, J.E. 1992, \apj, 393, 84
\bibitem[Barnes (1999)]{Barnes} Barnes, J.E. 1999, in ASP
 Conf. Ser. 187, The Evolution of Galaxies on Cosmological Timescales,
 ed. J.E. Beckman \& T.J. Mahoney (San Francisco: ASP), 293
\bibitem[Blumenthal et al.~(1984)]{BFPR} Blumenthal, G.R., Faber,
 S.M., Primack, J.R., \& Rees, M. 1984, Nature, 311, 517
\bibitem[Blumenthal et al.~(1986)]{BFFP} Blumenthal, G.R., Faber,
 S.M., Flores, R., \& Primack, J.R. 1986, \apj, 301, 27
\bibitem[Bond et al.~(1991)]{BondEPS}Bond, J.R., Cole, S., Efstathiou,
 G., \& Kaiser, N. 1991, \apj, 379, 440
\bibitem[Bower (1991)]{Bower}Bower, R. 1991, MNRAS, 248, 232
\bibitem[Bryan \& Norman (1998)]{BryanNorman}
 Bryan, G.L., \& Norman, M.L. 1998, \apj, 495, 80
\bibitem[Bullock et al.~(2001b)]{DTH}
 Bullock, J.S., Dekel, A., Kolatt, Kravtsov A.V., Klypin, A.A., 
 Porciani, C., \& Primack, J.R., 2001b, ApJ, 555, 240
\bibitem[Bullock et al.~(2001a)]{Bullock}
 Bullock, J.S., Kolatt, T.S., Sigad, Y., Somerville, R.S., Kravtsov
 A.V., Klypin, A.A., Primack, J.R., \& Dekel, A. 2001a, MNRAS, 321, 559.
\bibitem[Cohen (2000)]{Cohen} Cohen, J.G. 2000, \aj, 119, 162
\bibitem[Cole \& Lacey (1996)]{Cole}
 Cole, S., \& Lacey, C. 1996, MNRAS, 281, 716
\bibitem[Colin et al.~(1999)]{Pedro}
 Colin, P., Klypin, A.A., \& Kravtsov A.V, 2000, ApJ, 539, 561
\bibitem[C\^ot\'e et al.~(2001)]{Cote} C\^ot\'e, P., et al. 2001,
 \apj, 559, 828
\bibitem[Dekel et al.~(2001)]{Dekel}
 Dekel, A., Bullock, J.S., Porciani, C., Kravtsov, A.V.,
 Kolatt, T.S., Klypin, A.A., \& Primack, J.R., 2001, in ASP Conf. 
 Ser. 230, Galaxy Disks and Disk Galaxies (San Francisco: ASP),
 eds. J.G. Funes, S.J., and E.M. Corsini, astro-ph/0011002
\bibitem[Doroshkevich (1970)]{Dorosh}
 Doroshkevich, A.G. 1970, Astrofizika, 6, 581
\bibitem[Fall \& Efstathiou (1980)]{FallEf} Fall, S.M., \& Efstathiou,
 G. 1980, MNRAS, 193, 189
\bibitem[Flores et al.~(1993)]{FPBF} Flores, R., Primack, J.R.,
 Blumenthal, G.R., \& Faber, S.M. 1993, \apj, 412, 443
\bibitem[Franx, Illingworth \& Heckman (1989)]{Franx89}
 Franx, M., Illingworth, G.D., \& Heckman, T. 1989, \apj, 344, 613
\bibitem[Gamow (1952)]{Gamow}
 Gamow, G., 1952, Phys. Rev, 86, 251
\bibitem[Gardner (2001)]{Gardner} Gardner, J.P., 2001, ApJ, 557, 616
\bibitem[Hernquist (1993)]{Hern93} Hernquist, L. 1993, \apj, 409, 548
\bibitem[Hoyle (1949)]{Hoy}
 Hoyle, F., 1949, {\em Problems of Cosmical Aerodynamics}
 Dayton: Central Air Documents Office.
\bibitem[Hui et al.~(1995)]{Hui95} Hui, Xiaohui, Ford, H.C., Freeman,
 K.C., \& Dopita, M.A. 1995, \apj, 449, 592
\bibitem[Kissler-Patig \& Gerhardt (1998)]{KPGebhardt} Kissler-Patig,
 M., \& Gebhardt, K. 1998, AJ, 116, 2237
\bibitem[Klypin \& Holtzman (1997)]{KlypinHoltzman} Klypin, A.A., Holtzman, J., 1997, astro-ph/9712281
\bibitem[Klypin et al.~(2001)]{KKBP00} Klypin, A.A., Kravtsov, A.V.,
Bullock, J.S., \& Primack, J.R. 2001, ApJ, 554, 903 
\bibitem[Kravtsov et al.~(1997)]{ART}
 Kravtsov, A.V., Klypin, A.A., \& Khokhlov, A.M., 1997, ApJS, 111, 73 
\bibitem[Lacey \& Cole (1993)]{LaceyCole}Lacey, C., \& Cole, S. 1993,
 MNRAS, 263, 627
\bibitem[Lee \& Pen(2000)]{Lee1} Lee, J., \& Pen, U. 2000, \apjl, 532, L5 
\bibitem[Lee \& Pen(2001)]{Lee2}
 Lee, J., Pen, \& U., 2001, \apj, 555, 106
\bibitem[Lemson \& Kauffmann(1999)]{LemsonKauffmann} Lemson, G.,
 Kauffmann, G. 1999, \mnras, 302, 111 
\bibitem[Maller, Dekel, \& Somerville (2002)]{Maller} Maller, A.H.,
 Dekel, A., \& Somerville, R.S. 2002, \mnras, 329, 423
\bibitem[Mendez et al.~(2001)]{Mendez01} Mendez, R.H., et al. 2001,
 \apj, 563, 135
\bibitem[Mo, Mao, \& White (1998)]{MMW98} Mo, H.-J., Mao, S., \& White,
 S.D.M. 1998, \apj, 297, L71; MNRAS, 295, 319
\bibitem[Navarro, Frenk, \& White (1996)]{NFW96} Navarro, J.F., Frenk,
 C.S., \& White, S.D.M. 1996, \apj, 462, 563
\bibitem[Navarro, Frenk, \& White (1997)]{NFW97}
 Navarro, J.F., Frenk, C.S., \& White, S.D.M. 1997, ApJ, 490, 493
\bibitem[Navarro \& Steinmetz (1997)]{ns97}
 Navarro, J.F., \& Steinmetz, M. 1997,  ApJ, 478, 13
\bibitem[Padmanabhan (1993)]{Pad}
 Padmanabhan, T., 1993, {\em Structure formation in the
 universe}, Cambridge: Cambridge Univ. Press
\bibitem[Peebles (1969)]{Peebles} Peebles, P.J.E. 1969, ApJ, 155, 393
\bibitem[Pen, Lee, \& Seljak(2000)]{PenLeeS2000} Pen, U., Lee, J., \&
 Seljak, U. 2000, \apj, 543, L107 
\bibitem[Porciani et al.~(2002)]{Porciani2001}
 Porciani, C., Dekel, A., \& Hoffman, Y. 2002,  \mnras, 332, 325
\bibitem[Press \& Schechter (1974)]{Press}
 Press, W.H., \& Schechter, P. 1974, ApJ, 187,425
\bibitem[Ryden (1988)]{Ryden}
 Ryden, B.S. 1988, ApJ, 329, 589
\bibitem[Somerville \& Kolatt (1999)]{SomKol}
 Somerville, R.S., \& Kolatt, T.S. 1999, MNRAS, 305, 1
\bibitem[Somerville \& Primack (1999)]{SP99}Somerville, R.S., \& Primack,
 J.R. 1999, MNRAS, 310, 1087
\bibitem[Somerville et al.~(2000)]{SLKD} Somerville, R.S., Lemson,
 L.G., Kolatt, T.S., \& Dekel, A. 2000, MNRAS, 316, 479
\bibitem[Steinmetz \& Bartelmann (1995)]{Steinmetz}
 Steinmetz, M., \& Bartelmann, M. 1995, MNRAS, 272, 570
\bibitem[Sugerman et al.~(2000)]{Suger}
 Sugerman, B., Summers, F.J., Kamionkowski, M., 2000, MNRAS, 311,762
\bibitem[van den Bosch (1998)]{Bosch} van den Bosch, F. 1998, ApJ, 507, 601
\bibitem[van den Bosch, Burkert, \& Swaters (2001)]
{Bosch2} van den Bosch, F.C., Burkert A., \& Swaters R.A. 2001, 
MNRAS, 326, 1205
\bibitem[Walker, Mihos, \& Hernquist (1996)]{Walker} Walker, I.R.,
 Mihos, J.C., \& Hernquist, L. 1996, \apj, 460, 121
\bibitem[Warren et al.~(1992)]{Warren}
 Warren, M.S., Quinn, P.J., Salmon, J.K., \& Zurek, W.H. 1992, ApJ, 399, 405
\bibitem[Wechsler (2001)]{Wechsler} Wechsler, R.H. 2001, PhD
 Dissertation, University of California, Santa Cruz
\bibitem[Wechsler et al.~(2002)]{Wechsler02} Wechsler, R.H., Bullock, J.S.,
  Primack, J.R., Kravtsov, A.V., \& Dekel, A. 2002, ApJ, 568, 52
\bibitem[Weil \& Hernquist (1996)]{WeilHern} Weil, M.L., \& Hernquist,
 L. 1996, \apj, 460, 101
\bibitem[Wieszacker (1951)]{Weiz}
 Wieszacker, C.F. von, 1951, ApJ, 114, 165
\bibitem[White (1984)]{White84}
 White, S.D.M. 1984, ApJ, 286, 38

\end{thebibliography}
\end{document}